\documentclass{aastex}          
\usepackage{spr-astr-addons}    

\begin{document}
%
\title{VLBI observation of giant radio galaxy J1313+696 at 2.3/8.4 GHz}

\shorttitle{VLBI observation of J1313+696 at 2.3/8.4 GHz}
\shortauthors{X. Liu \& J. Liu}

\author{Xiang Liu\altaffilmark{1}} \and \author{Jun Liu\altaffilmark{1,2}}


\email{liux@uao.ac.cn}

\altaffiltext{1}{National Astronomical Observatories/Urumqi
observatory, CAS, 40-5 South Beijing Rd, Urumqi 830011, PR China}

\altaffiltext{2}{Graduate University of the Chinese Academy of
Sciences, Beijing 100049, PR China}

\begin{abstract}

We report the result of VLBI observation of the giant radio galaxy
J1313+696 (4C +69.15) at 2.3/8.4 GHz, only the core component of
the giant radio galaxy was detected in the VLBI observation at the
dual frequencies. The result shows a steep spectrum core with
$\alpha=-0.82$ ($S \propto \nu^{\alpha}$) between 2.3 GHz and 8.4
GHz. The steep spectrum core may be a sign of renewed activity.
Considering also the upper limit flux density of 2.0 mJy at 0.6
GHz from Konar et al. 2004 the core has a GHz-peaked spectrum,
implying that the core is compact and absorbed. Further high
resolution VLBI observations are needed to identify if the steep
spectrum core is consisting of a core and steep spectrum jet.

\end{abstract}


\section{Introduction}

It's known that the large radio sources are classified into two
types i.e. FRI and FRII-type according to the morphology of the
radio sources. Among the large radio sources, the radio galaxies
whose lobes span a (projected) distance of above 1 Mpc are called
giant radio galaxies (GRGs), and the majority of them are FRII
sources (Schoenmakers et al. 2001). A small number of the large
extended sources consisting of a pair of double lobes have been
called double-double radio galaxies (DDRGs), and the observed
structures of DDRGs suggest recurrent or interrupted central
activity as the origin of these sources (Schoenmakers et al.
2000a; Saikia et al. 2006). The large radio sources have been
observed with telescopes or arrays at relatively low resolution
but not well observed with the very long baseline interferometry
(VLBI). The VLBI observation is able to identify the core of the
large radio sources at high resolution. We have included the GRG
J1313+696 in our sample of VLBI observation with the European VLBI
Network (EVN) at dual frequency 2.3/8.4 GHz simultaneously.

\section{Observation and data reduction}

The VLBI observation was carried out on 2002 June 4 at 2.3/8.4 GHz
using the Mark4 recording system with a bandwidth of 16 MHz in
right circular polarization. The EVN antennae in this experiment
were Effelsberg, Wettzel, Medicina, Noto, Matera, Onsala, Yebes,
Urumqi and Shanghai. All station gives useful fringes. The
snapshot observations of 12 target sources in a total of 24 hours
of observing time have been done (Xiang et al. 2005). The source
J1313+696 was included in the source sample and observed in total
of about 1.5 hours. The calibrator source OQ208 was used in the
observation. The data correlation was completed at the Joint
Institute of VLBI in Europe (JIVE) in January 2003.

The Astronomical Image Processing System (AIPS) has been used for
editing, a-priori calibration, fringe-fitting, self-calibration
and imaging. The ANTAB file was checked and corrected according to
the EVN status table, station feedback and the observation
log-files. The AIPS task `UVCRS' was used to correct the estimated
antenna gains of Wettzel and Matera antenna before the
fringe-fitting, since their antenna gains were not available in
the EVN status table. The errors of the flux densities were
estimated to be in uncertainty of 10\% according to the calibrator
source OQ208 in the final VLBI images.

\section{Result and discussion}

J1313+696 (B1312+698, DA340, 4C +69.15, z=0.106), is an FRII type
giant radio galaxy. In VLA observation at 1.4 GHz the emission
from the core to the lobes forms a continuous bridge in SE-NW
direction (Lara et al. 2001), the GMRT observation at 605 MHz
shows a similar structure (Konar et al. 2004). But at 4.9 GHz only
the core and the lobe extremes are detected with the VLA (Lara et
al. 2001), and the core position is consistent with the position
of the associated galaxy. The source was not classified as a DDRG
according to the definition of Schoenmakers et al. (2000a) which
an inner pair of edge-brightened lobes should be detected.

\begin{figure}
     \includegraphics[width=6cm]{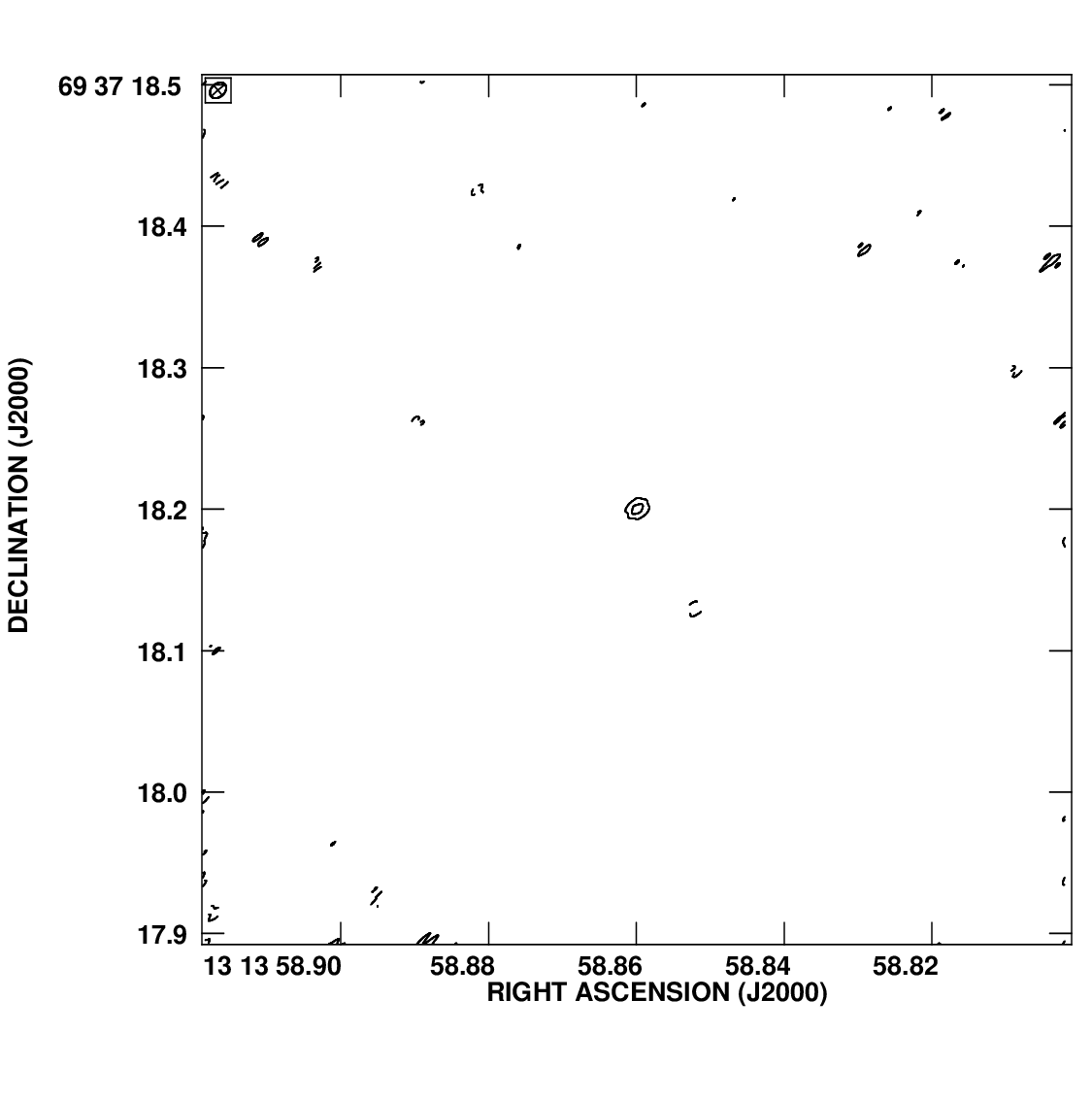}
     \caption{VLBI image of J1313+696 at 2.3 GHz, the restoring beam is
        $12.9\times9.0$ mas with PA $-49^{\circ}$, the peak is 7.7 mJy/beam, the first contour
        is 3 mJy/beam which is 3$\sigma$. The contour levels are increased by a factor of 2.}
      \label{fig1}
   \end{figure}

The VLBI images (Fig.~\ref{fig1}, Fig.~\ref{fig2}) of J1313+696 at
2.3/8.4 GHz show a point source. We checked the position of the
point source in the VLBI maps with the VLA images at 1.6 and 5 GHz
(Lara et al. 2001), confirmed that the VLBI point source is the
core of J1313+696 in the VLA images. We collected the core flux
densities of J1313+696 from our data and literature as shown in
Fig.~\ref{fig3}, the error bars are 1$\sigma$ (per beam) from our
data and the literature. The source flux densities in the VLBI
observation at 2.3 GHz and 8.4 GHz are 7.5$\pm$1.1 mJy and
2.6$\pm$0.2 mJy respectively. It is 10.2 mJy at 1.4 GHz and 3.8
mJy at 4.9 GHz in Lara et al. (2001), and 6 mJy at 2.7 GHz in
Saunders et al. (1987). It is 7 mJy at 1.4 GHz, 4.3 mJy at 4.9
GHz, and an upper limit of 2 mJy at 605 MHz in Konar et al.
(2004). The spectral index between 2.3 and 8.4 GHz in the VLBI
observation is $-0.82$ (we use $S \propto \nu^{\alpha}$). The
spectral index from the collected data above 1.4 GHz can be fitted
linearly (Fig.~\ref{fig3}) with $\alpha=-0.78\pm0.03$. It is
interesting that this value is consistent with the spectral
indices of the east, west lobe of J1313+696 (Schoenmakers et al.
2000b). However, because an upper limit 2 mJy at 605 MHz was
estimated by Konar et al. (2004) the integrated spectrum in
Fig.~\ref{fig3} can also be an inverted spectrum which peaked
around 1.4 GHz. We note that the 1.4 GHz core flux (10.2 mJy) from
Lara et al. (2001), might be overestimated due to contamination of
diffuse emission. As Lara et al. mentioned, for this core the
ratio peak/total $<0.8$. An ideal point source should have the
ratio peak/total $\sim$1. The core is sitting on top of a diffuse
emission, as is evident from the map in Lara et al.. Whereas, the
measurement of Konar et al. (2004) has less contamination as they
re-mapped the field with lower uv-cutoff to lose the diffuse flux
and get the core flux as correct as possible.

Konar et al. (2004) have discussed the steep-spectrum cores (SSCs,
$\alpha_{core} < -0.5$) of GRGs, in their sample at least 3 out of
17 sources show SSCs. In Lara et al. (2004) sample a compact radio
core was detected at 4.9 GHz with VLA in all sources, with an
average spectral index of $0.07\pm0.41$ for FRII and
$-0.24\pm0.52$ for FRI sources; hence FRIs seem to have more SSCs
than FRIIs according to the averaged spectral indices. However,
Lara et al. (2004) explained that the core spectral index in FRIs
suffers from more contamination from the steeper jet emission than
in FRIIs. In Konar et al. (2004) sample, 3 FRII-type sources have
been suggested to have SSCs besides J1313+696. They suggested that
SSC is preferentially occurred in giant radio galaxies, and SSC is
related to the interrupted activity of GRGs. Schoenmakers et al.
(2000b) found a polarized component of J1313+696 near but not at
the core position, they suggested it is a jet component.

The VLBI images of J1313+696 show a point source, not as expected
to show a core-jet, but reveal a steep spectrum core which may
imply a renewed jet in the core. Considering the upper limit flux
density at 605 MHz the core of the J1313+696 has a GHz-peaked
spectrum (GPS), indicating the core is very compact and absorbed
at lower frequency. We estimated an upper limit of the core size
at 8.4 GHz with the beam size of our VLBI map, it is 3.8 pc (in
$H_0=71~km s^{-1} Mpc^{-1}$, $\Omega_m$=0.27, and
$\Omega_{vac}$=0.73 cosmology). Another example is the DDRG
B1834+620, its core shows a steep spectrum at higher frequencies
with spectral index the same as that of lobes, and the core
spectrum is inverted at lower frequencies, showing a GPS shape
(Schoenmakers et al. 2000c). Konar et al. (2008) found a GPS core
in the GRG J1155+4029 with GMRT and VLA observations.

\section{Summary}

We present the first VLBI images of the giant radio galaxy
J1313+696 at 2.3/8.4 GHz, the VLBI result shows a steep spectrum
core of J1313+696 with $\alpha=-0.82$ between 2.3 GHz and 8.4 GHz.
The steep spectrum core may be a sign of renewed activity. The
core has a GHz-peaked spectrum, implying that the core is compact
and absorbed. Further high resolution VLBI observations are needed
to identify if the steep spectrum core is consisting of a core and
steep spectrum jet.

\begin{figure}
     \includegraphics[width=6cm]{liu_fig2.eps}
     \caption{VLBI image of J1313+696 at 8.4 GHz, the restoring beam is
        $2.1\times2.0$ mas with PA $28^{\circ}$, the peak is 2.7 mJy/beam, the first contour
        is 0.5 mJy/beam which is 3$\sigma$. The contour levels are increased by a factor of 2.}
      \label{fig2}
   \end{figure}

\begin{figure}
     \includegraphics[width=6cm]{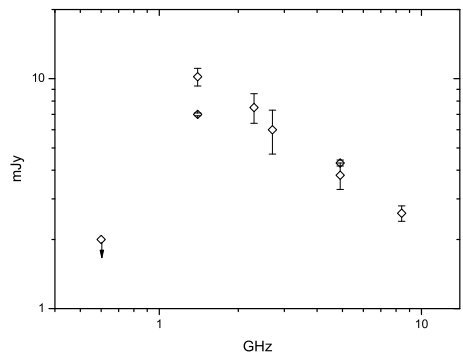}
     \caption{Spectrum of the core component of J1313+696 at 0.6, 1.4, 2.3, 2.7, 4.9, and 8.4 GHz. Data are from
     Konar et al. (2004), Lara et al. (2001), Saunders et al. (1987) and this observation.}
      \label{fig3}
   \end{figure}

\acknowledgments Valuable comment from the anonymous referee is
acknowledged. This work is supported by the National Natural
Science Foundation of China (NNSFC) under grant No.10773019 and
the 973 Program of China under grant No.2009CB824800.

\end{document}